\documentclass[11pt,a4paper]{article}
\pdfoutput=1
\usepackage{jheppub}
\usepackage{graphicx}
\usepackage{lipsum}
\usepackage[all]{hypcap}
\usepackage{soul}
\usepackage{xcolor}
\allowdisplaybreaks

\newcommand{\dis}[1]{\begin{equation}\begin{split}#1\end{split}\end{equation}}
\newcommand{\be}{\begin{equation}}
\newcommand{\ee}{\end{equation}}
\newcommand{\bea}{\begin{eqnarray}}
\newcommand{\eea}{\end{eqnarray}}
\newcommand{\beq}{\begin{equation}}
\newcommand{\eeq}{\end{equation}}
\newcommand{\comment}[1]{}

\def\hone{h_{(1,1)}}
\def\htwo{h_{(2,2)}}

\def\tr{{\rm tr}}
\def\c{{\hat{c}}}
\def\d{{\rm d}}
\newcommand{\Mp}{M_{\rm P}}

\def\V{\hat{V}}

\def\A{{\cal A}}
\def\M{{\cal M}}

\def\R{{\cal R}}
\def\S{{\cal S}}
\def\T{{\cal T}}
\def\X{{\cal X}}
\def\ls{\ell_s}
\def\lh{\ell_{11}}
\newcommand{\aYM}{\alpha_{\rm YM}}

\begin{document}

\begin{flushright}
\end{flushright}

{\raggedleft PNUTP-19-A12\\}

\title{Axion clockworks from heterotic M-theory:\\
the QCD-axion and its ultra-light companion }

\affiliation[a]{Department of Physics, Pusan National University, Busan 46241, Korea}
\affiliation[b]{Bethe Center for Theoretical Physics and Physikalisches Institut 
der Universit\"at Bonn,\\Nussallee 12, 53115 Bonn, Germany}


\affiliation[c]{Institute of Theoretical Physics, Faculty of Physics, University of Warsaw,\\ ul.~Pasteura 5, 02-093 Warsaw, Poland}
 
\author[a]{Sang Hui Im,}
\author[b]{Hans Peter Nilles,}
\author[c]{Marek Olechowski}

\emailAdd{imsanghui@pusan.ac.kr}
\emailAdd{nilles@th.physik.uni-bonn.de}
\emailAdd{Marek.Olechowski@fuw.edu.pl}

\abstract{
A previously discussed clockwork mechanism within heterotic M-theory is applied
to its axion landscape. We identify a unique candidate for a QCD-axion with a decay
constant in the preferred ``axion window" around $10^{11}\,$GeV. It is accompanied 
by at least one ultra-light axion that couples predominantly to hidden sector gauge groups.
}

\maketitle

\section{Introduction}

Axions have been considered in various areas of particle
physics and cosmology, such as the solution to the strong
CP-problem
\cite{Peccei:1977hh},
the origin of inflation
\cite{Freese:1990rb}
and their contribution to dark matter and dark energy
of the universe. These applications are typically connected
to peculiar relations (hierarchies) of the mass scales under
consideration. The QCD-axion as a solution of the strong
CP-problem has to have a decay constant in the so-called
``axion window" between $10^9$ and $10^{12}$ GeV
\cite{Kim:1986ax}
much larger than the QCD scale\footnote{Here we will assume the standard axion cosmology
without consideration of late time entropy injection \cite{Steinhardt:1983ia, Kawasaki:1995vt} or stronger QCD phase in the early universe \cite{Dvali:1995ce, Banks:1996ea, Co:2018phi} which may relieve
the upper bound $10^{12}$ GeV on the QCD axion decay constant.}; 
axionic inflation seems
to require trans-Planckian excursions of the inflaton
field \cite{Freese:2014nla, Marsh:2015xka}. These scales could be chosen ad hoc, but in a
consistent ultraviolet (UV) complete theory one would
like to understand the origin of these scales and
hierarchies as well as their relation to physical mass
scales like the Planck mass, the weak (TeV) scale and
the QCD scale.

A specific mechanism to create a hierarchy of axionic
scales has been proposed in the framework of axionic
(natural) inflation to create trans-Planckian axion
decay constants via the alignment of (two) axions
\cite{Kim:2004rp,Kappl:2014lra},
each with sub-Planckian decay constant. This alignment
mechanism has been generalised to the multi-axion case
\cite{Choi:2014rja}
that allows the creation of an even
larger hierarchy from moderate values of the
fundamental parameters. Such a multi-axion approach
has been used to create a large scale for the
QCD axion (within the ``axion window") starting from
weak scale physics in the TeV-range
\cite{Higaki:2015jag}.

A new challenge towards the creation of large
hierarchies appears with the proposal of the so-called
``relaxion mechanism" for a dynamical creation of
the weak scale from the Planck scale
\cite{Graham:2015cka}.
Again, the mechanism of axion alignment was applied
successfully
\cite{Choi:2015fiu}.
It was subsequently pointed out that there is a similarity
of the  aligned axions and a ``clockwork" mechanism
\cite{Kaplan:2015fuy}.
As such, the multi-axion systems discussed here resemble
approaches with accidental continuous symmetries (accions)
through discrete symmetries
\cite{Choi:2009jt}
or the consideration of deconstructed extra dimensions
\cite{ArkaniHamed:2001ca,Hill:2000mu}.
A next step considered the continuum limit
$N\to\infty$ of aligned $N$-axion systems
\cite{Giudice:2016yja}.
Discrete and continuous clockworks share some similarities
but differ in the detailed interpretation as discussed
in refs.
\cite{Craig:2017cda,Giudice:2017suc,Choi:2017ncj}.
While all these applications are very promising for the
creation of hierarchies, there still remains the question
of the embedding in a consistent ultraviolet completion.
In previous work
\cite{Im:2018dum}
we have considered the continuous clockwork mechanism in the
framework of heterotic M-theory of Ho\v rava and Witten
\cite{Horava:1996ma}.\footnote{As discussed in \cite{Choi:2017ncj},
the notion of continuous clockwork incorporates Randall-Sundrum (RS) model \cite{Randall:1999ee} and large
extra dimension (LED) \cite{ArkaniHamed:1998rs}
as well as more general warped extra dimensional models, such as linear dilaton model \cite{Antoniadis:2001sw}. 
 The extra dimensions of heterotic M-theory \cite{Lukas:1998tt} can be described in terms of 
 continuous clockwork as well \cite{Im:2018dum}.}
For certain specific clockwork parameters we could
accomodate a large hierarchy between the Planck scale and
the weak (TeV) scale.

In the present paper we investigate the options for the
QCD-axion as a solution to the strong CP-problem within a
heterotic M-theory clockwork scheme, i.e.~we explore the possibilities to
obtain an axion decay constant within the allowed
``axion window", extending previous discussions of QCD-axions
embedded in string theory
\cite{Svrcek:2006yi}.
Within the framework of the (weakly coupled) heterotic
string it turns out to be very difficult to arrange for
a large hierarchy between the Planck scale and the
axion scale because these scales are strictly related
through the value of the gauge coupling constants.
This is different in heterotic M-theory (strongly coupled
case) because of the appearance of warped extra dimensions.
The gauge coupling constants on the visible and hidden
brane are controlled by the volume of the compact
six-dimensional space (Calabi-Yau (CY) manifold) on the
respective brane. A hierarchy of scales could emerge in
the situation of particular non-standard embeddings
where the volume of the hidden brane is larger than that
of the visible brane. In our previous work
\cite{Im:2018dum}
we have discussed this clockwork mechanism towards 
a solution of the hierarchy problem between the weak
scale and the Planck scale. In the present paper we
explore the mechanism for the explanation of the
(more moderate) hierarchy between the Planck scale
and the QCD-axion decay constant. We illustrate the
mechanism in the simplest case of two axions known
as the model-independent (MI) and the model-dependent
(MD) axion
\cite{Witten:1984dg}.
This corresponds to the case of one K\" ahler modulus
of the Calabi-Yau manifold ($\hone=1$) and the
clockwork parameter $\hat c^2=6$
\cite{Im:2018dum}.
More general cases will be relegated to the
appendix.

Our analysis clarifies that there is no possibility for an acceptable QCD-axion
in weakly coupled heterotic string theory: the axion decay constants will always
be of order of the string scale and thus far above the ``axion window".
The case of heterotic M-theory, however,
looks more promising. The continuous clockwork mechanism discussed in ref.
\cite{Im:2018dum} allows a lowering of the relevant axion decay constant.
The results of the present paper can be summarised as
follows:
\begin{itemize}

\item There are various possibilities to lower axion decay constants to match the
``axion window".

\item Among them there is, however, only a unique candidate for a QCD axion
where $M_{11}$ (the fundamental mass scale of M-theory) is lowered to a value
not too far from that axion window. In fact, the axion decay constant $f_a$
of the QCD axion will be an order
of magnitude smaller than $M_{11}$.

\item This QCD axion  is accompanied by at least one (hidden) ultra-light axion (ULA).
Its decay constant is smaller than that of the QCD axion and it is essentially decoupled from the
observable sector. Properties of such ULA(s) should be investigated thoroughly.

\end{itemize}
The paper is structured as follows. In section \ref{sec_axions_def} we shall present the axion candidates,
separately for the weakly coupled heterotic string and heterotic M-theory. Section \ref{sec_axions_mixing}
discusses the mixing between the various axion candidates and points out specific
differences between the cases of string and M-theory. In section \ref{sec_decay_const} we shall analyse
the axion decay constants for the two separate cases. Section \ref{sec_QCD_axion} is devoted to the
analysis of candidates for a QCD axion. There we shall present the unique axion
solution to the strong CP-problem. Section \ref{sec_conclusions} will be devoted to 
 conclusions and outlook. Some technical details and discussion of Kaluza-Klein (KK) axions  that originate from the clockwork mechanism  will be given in the appendices.

\section{Axion Candidates} 
\label{sec_axions_def}


In this section, we will review the axion candidates under consideration. We shall identify
and compare the candidates in the standard 10-dimensional 
heterotic string theory (weakly coupled
case) with those in 11-dimensional heterotic M-theory (strongly coupled case).

\subsection{Weakly coupled heterotic string}

The part of the low energy action of the heterotic string theory relevant for our discussion can be written as\footnote{We adopt the conventions used in \cite{Svrcek:2006yi}.}
\begin{align}
  \S_{10}
  =
  \frac{1}{2\kappa_{10}^2} \int_{{\cal M}^{10}}
  \left(\star\R 
  - \frac{\ls^4}{2} H \wedge \star H
- \frac{\alpha'}{4} {\rm tr} F \wedge \star F 
\right) 
-\int_{{\cal M}^{10}}  B \wedge X_8+\ldots
  \label{10Daction}
\end{align}
where 
$\kappa_{10}^2 = e^{2\phi} (\alpha')^4/8$,  the string tension $\alpha'$ and length $\ls$ are related by $\alpha' = (\ls/2\pi)^2$,  
$H$ is the field strength of the 2-form field $B$,  $F$ is the
$E_8 \times E_8$ gauge field strength. The first term is understood as the classical action of heterotic string theory, while the second integral in the rhs of the above action is necessary for the Green-Schwarz (GS) anomaly cancellation mechanism \cite{Green:1984sg} with the 8-form constructed out of $F$ and $\R$ 2-forms: 
$X_8 = c_1 \,\tr F^4 + c_2\,(\tr F^2)^2 + c_3\,\tr F^2\,\tr\R^2  + \cdots$. It could be interpreted as a 
1-loop quantum correction to the classical theory.

The field strength $H$ has to satisfy the modified Bianchi identity
\begin{equation}
\label{Bianchi_H}
dH = -\frac{1}{16\pi^2} \left( {\rm tr} F \wedge F - {\rm tr} \R \wedge \R \right)
\end{equation}
for the theory to be supersymmetric and anomaly free. This leads to the following expression for $H$
\begin{equation}
\label{H_def}
H = dB + \omega_3\,,
\end{equation}
where
\begin{equation}
\label{omega3def}
\omega_3 \equiv -\frac{1}{16\pi^2} \left( \omega_{YM} - \omega_{L}\right)
\end{equation}
with
\begin{equation}
\label{cs_form}
\omega_{YM} \equiv \tr\left(AF - \frac{2}{3} A^3 \right)
\,,\qquad 
\omega_{L} \equiv \tr \left(\omega \R - \frac{2}{3} \omega^3 \right)\,,
\end{equation}
which are the Yang-Mills and Lorentz Chern-Simons 3-forms, respectively.

We are interested mainly in the axion fields in ${\cal{N}}=1$ SUSY theory in 4D, so in order to get their definitions and interactions we compactify the 10D theory described by the action \eqref{10Daction} on a compact 6D CY manifold ${\cal{X}}^{6}$. Thus, we assume $\M^{10}=\M^4\times\X^6$ with $\M^4$ being the 4D Minkowski space-time. 
For the reduction procedure, we expand $B$ and $\omega_3$ forms in harmonic 
$(1,1)$-forms $\omega^{(1,1)}_i \in H^{(1,1)}(X) \, (i=1, \dots, \hone)$ on $\X^6$:
\begin{eqnarray}
\label{B}
B &=& \frac{1}{2\pi} \omega^{(1,1)}_i \, b^i(x) +\cdots \,,
\\
\label{omega3}
\omega_3 &=& \frac{1}{2\pi} \omega^{(1,1)}_i \wedge q^i A(x) +\cdots \,,
\end{eqnarray}
where $b^i(x)$ and $A(x)$ are 4D fields and factors of $1/2\pi$ are introduced to obtain $2\pi$-periodicity of the resulting axions and $q^i$ are dimensionless numbers depending on the gauge bundles over $\X^6$. The important part of the 8-form $X_8$ from the Green-Schwarz term is given by 
\begin{equation}
\label{X8}
X_8 = \omega^{(3,3)} \wedge q F(x) + \frac{1}{8\pi}  \left(\tr_1 F \wedge F - \frac{1}{2} \tr \R \wedge \R\right) \wedge\left(\tr_1 F \wedge F - \tr_2 F \wedge F\right) + \cdots \,,
\end{equation} 
where $\omega^{(3,3)}$ is the harmonic $(3,3)$ form on the CY space $\X^6$, $q$ is another dimensionless number determined from the gauge bundles over $\X^6$, and $\tr_j$ is trace in the $j$-th $E_8$ gauge group.

Using eqs.~\eqref{B}, \eqref{omega3} and \eqref{X8} in the action \eqref{10Daction}, the relevant part of the resultant 4D action may be written in the form
\begin{equation}
\label{S4a}
\S_{4, \textrm{axion}}
  = \S_{4, \textrm{MI}} + \S_{4, \textrm{MD}}
\end{equation}
with the first term in the rhs given by
\begin{equation}
\label{S4MI}
\S_{4, \textrm{MI}} 
= 
-\frac{\ls^4 V_{X}}{2\kappa_{10}^2}\int_{{\cal M}^{4}} 
 \frac{1}{2} H \wedge \star H -\int_{{\cal M}^4} q B \wedge F \, .
\end{equation}
This part of the action describes dynamics of the 3-form $H$ which in 4D may be dualized to a scalar. A way to dualize $H_{\mu\nu\rho}$ 
\footnote{When referring to components we use the following conventions: indices from the middle of the Greek alphabet are tangent to Minkowski space ${\cal{M}}^4$, $\mu,\nu,\ldots=0,1,2,3$; indices from the middle of the Latin alphabet are tangent to CY space ${\cal{X}}^6$, $m, n,\ldots=5,6,7,8,9,10$; indices from the beginning of the Greek alphabet are tangent to ${\cal{M}}^4\times S^1$,  $\alpha,\beta,\ldots=0,1,2,3,11$.
\label{convention}
}
is to enforce the Bianchi identity \eqref{Bianchi_H} by adding to \eqref{S4MI} a term with a 1-form Lagrange multiplier. Then $H$ may be eliminated using its equation of motion first and action \eqref{S4MI} may be rewritten in the form
\begin{equation}
\label{S4MIa}
\S_{4, \textrm{MI}} 
= 
-\frac{2\kappa_{10}^2}{V_X \ls^4} \int_{{\cal M}^{4}} \frac{1}{2} \left(da + q A\right) \wedge \star \left(da + q A\right) 
+ \frac{1}{16\pi^2} \int_{{\cal M}^{4}}  a \left(\tr_1 F \wedge F + \tr_2 F \wedge F\right)\,,
\end{equation}
where $a$ is a scalar field to be identified as an axion. It couples the same way to both visible and hidden $E_8$ sectors and does not depend on the details of the compactification from 10D to 4D. Thus, it is called the model independent (MI) axion. Loosely speaking, it corresponds to the pseudoscalar $B_{\mu\nu}$ and couples directly to the instantons in the classical action.

The second term in the rhs of \eqref{S4a} reads
\begin{equation}
\label{S4MD}
\S_{4, \textrm{MD}} 
= 
-\frac{\ls^4}{16\pi^2\kappa_{10}^2}\,G_{ij} \int_{{\cal M}^{4}} (d b^i + q^i A) \wedge \star (d b^j + q^j A) 
 +\frac{n_i}{16\pi^2}\int_{{\cal M}^4}   b^i \left(\tr_1 F \wedge F - \tr_2 F \wedge F\right),
\end{equation}
where
\begin{equation}
\label{Gij}
G_{ij} 
=  \int_{\X^6} \omega_i^{(1,1)} \wedge \star \omega_j^{(1,1)} \,, \end{equation}
\begin{equation}
\label{ni}
n_i 
= -
\frac{1}{16\pi^2} \int_{\X^6} \omega_i^{(1,1)} \wedge \left(\tr_1 F \wedge F - \frac{1}{2} \tr \R \wedge \R\right)\,. 
\end{equation}
Action \eqref{S4MD} describes $\hone$ axions $b^i$. They are called model dependent (MD) axions because, contrary to the MI axion $a$, their interactions depend on the details of the compactification via quantities $G_{ij}$ and $n_i$.  Loosely speaking they
can be identified with the pseudoscalars $B_{mn}$, with indices $m,n$ in the compact space.
They  have opposite sign couplings to visible and hidden $E_8$ gauge groups via
the Green-Schwarz polynomial.
Note that fields $a$ and $b^i$ are normalized to be $2\pi$-periodic. 
In the 4D supergravity language axions are pseudoscalar components of chiral superfields. The MI axion belongs to the dilaton field $\S$ while the MD axions are parts of the moduli $\T^i$ (the overall modulus $\T$ in the case of $\hone=1$). At the classical level, these axions do not mix.
Quantum corrections lead to a mixing of the axions as will be discussed in section 3.

In the above discussions, we have ignored $q A$ and $q^i A$ for a while. In fact, some of the charges $q$ and $q^i$ do not vanish if there exist anomalous $U(1)$ gauge symmetries in $E_8^{(1)} \times E_8^{(2)}$ under which the corresponding axions transform non-linearly.
Such axions may be gauged away becoming longitudinal components of the gauge fields. 
Yet one can have remnant global $U(1)_{\rm PQ}$ symmetries in matter sectors to deal with the strong CP problem \cite{Witten:1984dg, Barr:1985hk, Svrcek:2006yi}.
In the next section, we will see that a similar structure appears for axions from the strongly coupled heterotic string,
but the associated $U(1)$ gauge symmetries are not directly related to 
 $E_8^{(1)} \times E_8^{(2)}$. They instead come from the bulk gauge invariance of the 3-form $C$ of the 11D supergravity. 
This will give a clockwork structure for the axions in the strongly coupled theory.

\subsection{Heterotic M-theory}

The strongly coupled $E_8 \times E_8$ heterotic string theory (heterotic M-theory) can be described in terms of 11D supergravity with the bosonic part of the action given by
\begin{align}
  \S_{11}
  =
  &\,\,
  \frac{1}{2\kappa^2_{11}} \int_{{\cal M}^{11}} 
  \left( \star\R - \frac{\lh^6}{2} G \wedge \star G
  -\frac{\lh^9}{6} C \wedge G \wedge G \right)
  \nonumber\\  &\,\,
  -\frac{1}{8\pi \kappa^2_{11}} \left(\frac{\kappa_{11}}{4\pi}\right)^{2/3}
  \sum_{j=1}^{2}
  \int_{{\cal M}^{10}_{(j)}} 
   {\rm tr} F_{(j)} \wedge \star F_{(j)}
\,,
  \label{11Daction}
\end{align}
where $\kappa^2_{11}$ is related to fundamental 11D mass and length scales as follows: $\kappa^2_{11}=M_{11}^{-9}$ and $\kappa^2_{11}=\lh^9/4\pi$. $G$ is the field strengths of the 3-form field $C$
\footnote{
{The normalization convention for $G$ and $C$ is adopted from \cite{Svrcek:2006yi}, and differs from that in \cite{Im:2018dum} 
by factor $\lh^3/\sqrt{2}$ and $\lh^3/6\sqrt{2}$, respectively.}
}, $F_{(j)}$ are the gauge field strengths of two $E_8^{(j)}\, (j=1,2)$ gauge groups living at the two 10D orbifold fixed hyperplanes $\M_{(j)}^{10}$.

In analogy to the weakly coupled heterotic string, the field strength $G$ has to satisfy a modified Bianchi identity
which in this case takes the form
\dis{ \label{Bianchi_G}
\d G =J^{(1)} \wedge \delta(x^{11})\, \d x^{11}  
+J^{(2)} \wedge  \delta(x^{11}-\pi r_{11}) \,\d x^{11}
}
with
\dis{
J^{(j)} \equiv {-\frac{1}{8\pi^2}}  \left( {\rm tr} F_{(j)} \wedge F_{(j)} -\frac{1}{2} {\rm tr} \R\wedge \R \right) = \d \omega_3^{(j)}\,,
}
where 
\dis{ \label{omega3i}
 \omega_3^{(j)} \equiv {-\frac{1}{8\pi^2}} \left(\omega_{YM}^{(j)} - \frac{1}{2}\omega_{L}\right)\,,
 }  
while $\omega_{YM}^{(j)}$ and $\omega_{L}$ are the Yang-Mills and Lorentz Chern-Simons 3-forms  as in (\ref{cs_form}). 
The above Bianchi identity is solved when $G$ is defined as \cite{Horava:1996ma, Conrad:1997ww} 
\dis{ \label{G_def}
G = \d C +\omega_4\,,
}
where
\dis{ \label{dw4}
 \omega_4 = \omega_3^{(1)} \wedge \delta(x^{11})\, \d x^{11}  
+\omega_3^{(2)} \wedge  \delta(x^{11}-\pi r_{11}) \,\d x^{11} \,.
}
The $G$-field has a background value given by \cite{Witten:1996mz, Horava:1996ma, Lukas:1998yy, Lukas:1998tt}
\dis{ \label{G_bg}
\langle G \rangle= n_i \, \omega^{i (2,2)} \epsilon(x^{11}) \,,
}
where numbers $n_i$ are defined in \eqref{ni} and $\omega^{i (2,2)}\, (i=1, \dots, \htwo=\hone)$ are harmonic $(2,2)$ forms on $\X^6$ which satisfy the condition $\int_{\X^6} \omega_i^{(1,1)} \wedge \omega^{j (2,2)} = \delta_i^j$.

In order to find the 4D effective axion we first construct a 5D effective theory by dimensional reduction of the 11D action \eqref{11Daction}. For this purpose we perform the following steps:
%
%
We keep in $C$ and $G$ all relevant terms i.e.~their background values and light 5D components:
\begin{align}
C &= \langle C \rangle + \omega_i^{(1,1)} \wedge \A^i(x^\alpha) 
+ C^{(5)}(x^\alpha) + \ldots\,,
\\
G &= \langle G \rangle + \d\left(\omega_i^{(1,1)} \wedge \A^i(x^\alpha)\right) 
+ G^{(5)}(x^\alpha) + \ldots\,,
\end{align}
where 1-form fields $\A^i(x)$ and tensors $C^{(5)}$ and $G^{(5)}$ are 5D fields (have all indices tangent to 5D space and are functions of 5D coordinates) while $\langle C \rangle$ may be obtained from eqs.~\eqref{G_def}-\eqref{G_bg}. We substitute such $C$ and $G$ into the 11D action \eqref{11Daction} and add to it a term with a 7-form Lagrange multiplier to implement the modified Bianchi identity \eqref{Bianchi_G}. Then, we integrate over the CY space. Finally, eliminating first the tensor fields $C^{(5)}$ and $G^{(5)}$ (using their equations of motion), we get the following relevant part of the 5D action
\begin{align}
\label{5D_axion} 
\S_{5, {\rm axion}}=&\, -2\pi \int_{\M^5} \left[
\frac{\lh^3}{2 V_X} 
\left( \d a + {n_i \A^i}\right) \wedge \star \left( \d a + {n_i \A^i}\right) 
+ \frac{1}{2\lh^3} G_{ij}\,\d\A^i \wedge \star \d\A^j 
\right]
\nonumber\\
 &\, + \int_{\M^5} \frac{1}{4\pi}\, a 
\left[
\left( {\rm tr} F_{(1)} \wedge F_{(1)} \right)\delta(x^{11}) 
+ \left( {\rm tr} F_{(2)} \wedge F_{(2)} \right) 
\delta(x^{11}-\pi r_{11})
\right]
\wedge\d x^{11} \,,
\end{align}
%
where kinetic function $G_{ij}$ was defined in \eqref{Gij} and
$a$ is a scalar related to the field dual to $G^{(5)}$ 
by the relation
\begin{equation}
\label{starG}
\star_{{}_5} G^{(5)} = -\frac{\ell_{11}^3}{V_X}\left(\d a + n_i\A^i\right)\,.
\end{equation}
The scalar $a$ couples the same way to both $E_8$ gauge groups in analogy to the MI axion in \eqref{S4MIa}. It is a priori not obvious whether the action \eqref{5D_axion}  describes any more axions. The only fields present in \eqref{5D_axion} (in addition to the scalar $a$ and the gauge fields strengths $F_{(j)}$) are $\hone$ 5D vectors $\A^i$. We will see later that the components $\A^i_{11}$, which are 4D scalars, will correspond to 4D axions. The first line in \eqref{5D_axion} shows that they necessarily mix with the scalar $a$. Thus, the mixing among these fields must be taken into account in order to identify precisely all axions in this case. Such mixing will be discussed in the next section.

The action \eqref{5D_axion} is somewhat similar to the action \eqref{S4MIa} describing the MI axion in the weakly coupled case. Here, however, the definition of the MI axion appears to be ambiguous. Contrary to the weakly coupled string case, dualization of the tensor field strength involves also other axion candidates $\A^i_{11}$ with  quantities depending on compactification details -- numbers $n_i$ defined in \eqref{ni} appear in relation \eqref{starG}.
However, comparing the weakly and strongly coupled heterotic theories, we notice that axion fields have similar origin in both cases. The field $a$ originates from dualization of the tensor field strength ($H$ in heterotic string theory and $G$ in heterotic M-theory) with no indices tangent to the CY space $\X^6$. The remaining axion fields are (parts of) coefficients in the expansion of the tensor field ($B$ in heterotic string theory and $C$ in heterotic M-theory) in terms of $\hone$ harmonic (1,1)-forms on $\X^6$. Their number and properties do depend on the details of the compactification. Thus, one could call them MD-like axions and refer to $a$ as the MI-like 
one.\footnote{
The field $a$ corresponds to $C_{\mu\nu,11}$ (while $C_{\mu\nu\rho}$  is responsible for the derivative of $a$ along the 11th dimension) in the strongly coupled case and to $B_{\mu\nu}$ in the weakly coupled one. The candidates for the remaining axions originate from $C_{mn,11}$ and $B_{mn}$, respectively.}

There are two differences between the weakly and strongly coupled theories due to the fact that in the latter case compactification on 
$\X^6$ leads to a 5D effective theory. One of them consists in different transformation properties of MD axion fields. In the strongly coupled case they are components of 5D vectors and become scalars only after further compactification to 4D (field $a$ is a scalar already in 5D). The second difference is quite obvious: all the relevant fields in the M-theory case depend on 5D coordinates and only (some of) their 4D zero modes play the role of 4D axions.

\section{Mixing}
\label{sec_axions_mixing}

\subsection{Weakly coupled heterotic string}

In the 4D supergravity language axions are imaginary (pseudoscalar) components of chiral superfields. The MI axion belongs to the dilaton filed $\S$ while the MD axions are parts of the moduli $\T^i$ (the overall modulus $\T$ in the case of $\hone=1$). While the axions are well-defined at the level of the classical action, they mix via quantum corrections connected to the Green-Schwarz anomaly cancellation terms. In the simple case of the standard embedding the observable sector gauge group is broken to $E_6$ and one obtains
\begin{equation}
f_6=\S+\epsilon\T\,,
\qquad
f_8=\S-\epsilon\T\,,
\end{equation}
where $f_6$ ($f_8$) is the gauge kinetic function of the observable (hidden) $E_6$ ($E_8$) and $\epsilon$ is a constant (fixed by the anomaly) 
\cite{Derendinger:1985cv,Choi:1985bz,Ibanez:1986xy}.
In the case of some simple compactification schemes (e.g.~orbifolds) this mixing can be calculated in explicit one-loop calculations of threshold corrections to the gauge kinetic functions
\cite{Dixon:1990pc,Stieberger:1998yi}.

The identification of  the physical QCD-axion will have to take into account this mixing. We shall not discuss this here in detail as we shall see later that it is impossible to obtain an axion decay constant within the ``axion window" in the weakly coupled heterotic string. Both decay constants $f_6$ and $f_8$ will be of order of the string scale and cannot be lowered in the desired way. This will be different in the strongly coupled case.

\subsection{Heterotic M-theory}

As explained in the previous section, the axions in the heterotic M-theory originate from the 4D zero modes of the 5D fields $a$ (MI-like)
and $\A^i_{11}$ (MD-like). 
We want to determine the mixing among them. However, it is obvious from \eqref{5D_axion} that these fields mix also with the remaining components 
of the 5D vectors $\A^i$. Thus, we have to eliminate this mixing first and this can be done by
a choice of the gauge.
The kinetic term for $\A^i$ present in \eqref{5D_axion} (obtained from that of the $G$ field strength in the 11D action) depends on compactification details because the gauge kinetic function $G_{ij}$, defined in \eqref{Gij}, is related to properties of (1,1)-forms on the internal CY space ${\X}^6$.
In this section we consider the simplest case
\footnote{
In Appendix \ref{app_h11=2}, we will work out a next simple example with $\hone = 2$. We expect that essential features should not be much different for bigger values of $\hone$.
} 
when there is only one such form (i.e.~$\hone=1$) so there is only one gauge field $\A\equiv\A^1$. In this case, it turns out that
\begin{equation} \label{G11}
G_{11} = 3 V_X^{1/3}.
\end{equation}
The mixing between $\A_\mu$ and $(\A_{11}, a)$ may be eliminated by choosing an appropriate gauge i.e.~by adding the following gauge fixing term \cite{Choi:2017ncj}
\begin{equation}
\label{Sgf}
S_{\rm g.f.}=-\int_{\M^5} \frac{\sqrt{-g}}{2g_\A^2} \V_X^{1/3} \left[g^{\alpha \beta} \partial_\alpha \A_\beta +g^{11,11}  \frac{\partial_{11} \chi}{\chi} \A_{11} + n_1 \V_X^{-4/3} g_\A^2 F_a^3 a \right]^2, 
\end{equation}
where
\begin{eqnarray}
g_\A^2 &=& \frac{\lh^3}{6\pi V_{X,0}^{1/3}}\, \\
F_a^3 &\equiv& \frac{2\pi \lh^3}{V_{X,0}}\,, \\
\chi  &\equiv& \sqrt{-g}\, g^{11,11} g^{\mu \nu} \frac{\eta_{\mu \nu}}{4} \V_X^{1/3}\,,
\end{eqnarray}
and $\V_X \equiv V_X/V_{X, 0}$ with $V_{X, 0}$ being the CY volume at our boundary $x^{11}=0$ (convention for indices as in footnote \ref{convention}). The 11D space-time is warped, so the background metric $g_{\alpha \beta}$ and the volume $V_X$ depend in general on the 11-th coordinate $x^{11}$.
Thus, the gauge choice depends on the background geometry. In our previous work \cite{Im:2018dum} we have shown that the 5D background geometry is given by
\dis{ \label{ds2J}
\d s^2 = \V_X^{-2/3} \left( e^{2k_1 x^{11}} \eta_{\mu\nu}\d x^\mu \d x^\nu + e^{2k_2 x^{11}} (\d x^{11})^2 \right). 
}
The prefactor $\V_X^{-2/3}$ is introduced to obtain the Einstein frame in {5D}. The exponents $k_1$ and $k_2$ may be written as
\dis{ \label{k1k2}
k_1 = - \hat{b}\, n\, M_{11}\,,
\qquad
k_2= \hat{c}^2 k_1 \, , 
}
where parameters $\hat{b}$ and $\hat{c}^2 > 0$ are determined by properties of the K\"ahler moduli, and $n$ is given by
\dis{
 n = -\frac{1}{V_{X,0}^{1/3}}\int_{\X^6} \omega \wedge \frac{1}{16\pi^2} \left(\tr F_{(1)} \wedge F_{(1)} - \frac{1}{2} \tr \R \wedge \R\right), 
}
with the K\"ahler form $\omega$ of $\X^6$. 
The sign of this number is important because it determines the direction of warping along the 11-th dimension. 
It is positive (negative $k_1$ and $k_2$) for the standard embedding while it can be negative (postive $k_1$ and $k_2$)
for non-standard embeddings. Only the latter case 
is of interest for our discussion as it
will give us the possibility of lowering the axion scale significantly.
For the present case of $\hone=1$ the above parameters read
\dis{
\hat{b} = \frac{\alpha_{\rm GUT}^{2/3}}{6 (4\pi)^{1/9}}\,,
\qquad
\hat{c}^2 = 6\,, \label{h11=1_bc}
}
and $n=n_1$ as defined in \eqref{ni} with $V_X\to V_{X,0}$. 
The volume modulus $\V_X$ is related to the dilaton $S$ of the
General Linear Dilaton (GLD) model \cite{Choi:2017ncj, Im:2018dum} which has the ``linear dilaton" background:
\dis{
\V_X = e^{\sqrt{2} S} = e^{k_2 y}\,. \label{h11=1_V}
}

After the gauge fixing \eqref{Sgf}, $\A_\mu$ is decoupled from $\A_{11}$ and $a$, and 
we are left with mixing between $\A_{11}$ and $a$ only. 
Diagonalizing it further, we obtain our final form for the 5D axion lagrangian: 
\begin{align}
S_{5, \rm axion} = - \int_{\M^5}  \d^5 x
&\left[\frac{1}{2} \eta^{\mu \nu} \partial_\mu \phi_L \partial_\nu \phi_L +\frac{1}{2} \eta^{\mu \nu} \partial_\mu \phi_R \partial_\nu \phi_R \right.
\nonumber\\
\label{cw_axion}
&\,\,+ \frac{1}{2} e^{2(k_1-k_2)x^{11}} \left[(\partial_{11} \phi_L + m_\A \phi_L)^2 + (\partial_{11} \phi_R - m_\A \phi_R)^2 \right] \\
&\,\,\left.+\frac{1}{4\pi} a \left( \,{\rm tr} F_{(1)} \widetilde{F}_{(1)} \delta(x^{11})  +  {\rm tr} F_{(2)} \widetilde{F}_{(2)} \delta(x^{11}-\pi r_{11}) \right)\right],
\nonumber
\end{align} 
where
\begin{equation}
m_\A = \sqrt{\left(\frac12 k_2-k_1\right)^2 + {n_1^2} g_\A^2 F_a^3}
\,= \frac12 k_2 + k_1\,. \label{mA}
\end{equation}
Note that, in this notation, only the MI-like field $a$ appears to couple to the gauge fields
$F_{(j)}\widetilde{F}_{(j)}$. The coupling of the MD-like axions is hidden in the mixing.
The 
orthogonal axion fields $\phi_L$ and $\phi_R$ are expressed in terms of the axionic fields $a$ and $\A_{11}$ via the relations
\begin{eqnarray}
\label{phiL}
\phi_L 
&=&
e^{-\left(\frac12 k_2-k_1\right) x^{11}}
\left(a\,F_a^{3/2} \,\cos\beta
-
\A_{11}\,g_\A^{-1}\, \sin\beta\right),  
\\
\label{phiR}
\phi_R
&=&
e^{-\left(\frac12 k_2-k_1\right) x^{11}}
\left(-a\,F_a^{3/2} \,\sin\beta
-
\A_{11}\,g_\A^{-1}\,\cos\beta\right), 
\end{eqnarray}
with the mixing angle $\beta$ satisfying the condition
\begin{equation} \label{mixing_angle}
\tan 2\beta = -g_\A F_a^{3/2} \frac{n_1}{\frac12k_2-k_1}
=\frac{\sqrt{2 k_1 k_2}}{\frac{1}{2} k_2-k_1} \,.
\end{equation}
For $\hat{c}^2 = 6\, (k_2 = 6k_1)$, the mixing angle $\beta$ comes out as
\dis{\beta = \frac{\pi}{6}\,. }
The profiles of these axion fields in the 11-th dimension will be important for the discussion of possible candidates for a QCD-axion that solve the strong CP-problem on the observable brane. This will be explicitely discussed in the next section.

From (\ref{cw_axion}), we see that the axions from heterotic M-theory are identified as ``continuous clockwork axions" as explicitly 
discussed in \cite{Craig:2017cda, Choi:2017ncj}. The bulk gauge boson mass $m_\A$ serves as the clockwork parameter for the zero mode coupling
hierarchy, while $k_1-k_2 \equiv p$ controls the clockwork gear (KK mode) masses \cite{Im:2018dum, Choi:2017ncj}.
An important difference from ``discrete clockwork axion" 
\cite{Choi:2015fiu, Kaplan:2015fuy} is that $\phi_L$ and $\phi_R$ couple to 4D instantons only in the
 following combination: 
\begin{equation}
a= F_a^{-3/2} \,e^{\left(\frac{k_2}{2}-k_1\right) x^{11}} \left( \phi_L \cos \beta - \phi_R \sin \beta \right).
\end{equation}
There is no such restriction in the discrete clockwork axion models. 
This is an example of the  limitations of this continuous clockwork axion compared to the discrete models
as pointed out in section 3.4 of \cite{Choi:2017ncj}.\footnote{More specifically, none of these continuous clockwork axions can have a short periodic ``wiggle" on top of
its cosine potential due to the presence of the other axion.}

\section{Decay constants and couplings}
\label{sec_decay_const}

Let us now turn to a discussion of the scales of the axion decay constants. 

\subsection{Weakly coupled heterotic string}

In the weakly coupled heterotic string the MI axion decay constant 
may be read off from the action \eqref{S4MIa} and equals
\begin{equation}
f_a = \sqrt{\frac{2\kappa_{10}^2}{V_X \ls^4}} = \frac{\alpha_{\rm YM}}{{2}\sqrt{2} \pi} \Mp \, ,
\end{equation}
where $\Mp$ is the 4D reduced Planck mass and $\alpha_{\rm YM}$ is the $E_8 \times E_8$ gauge coupling given by 
\begin{equation}
\Mp^2 = \frac{V_X}{\kappa_{10}^2}\,, 
\qquad 
\alpha_{\rm YM}^{-1} = \pi \alpha' \frac{V_X}{\kappa_{10}^2}\,. \label{weak_Mp}
\end{equation}
The MD axions $b^i$ mix for general non-diagonal $G_{ij}$ and their decay constants depend on the details of the model. A typical value 
of such constants may be estimated as
\begin{equation}
f_{b} 
\sim 
\sqrt{\frac{\ls^4}{8\pi^2\kappa_{10}^2}} \sqrt{\frac{V_X}{V_{C}^2}} 
= \frac{1}{2\pi f_a V_{C}} \gtrsim
\frac{\alpha_{\rm YM}}{2\sqrt{2} \pi} \Mp \, ,
\end{equation}
where $V_C$ measures an average volume of $(1,1)$ basis cycles $C_i$ on $\X^6$, and we have used $G_{ij} \sim V_X/V_C^2$ from 
(\ref{Gij}) because $\int_{C_i} \omega_i =1$. The inequality derives from the upper bound $V_C \lesssim \aYM^{-1} \ls^2$ as argued in \cite{Svrcek:2006yi}.
Therefore, both MI and MD axions have their decay constants around the GUT scale $\sim 10^{16}$ GeV.
This is due to the fact that the 
string scale is tied to the 4D Planck scale by the relation $\Mp^2 = (\pi \aYM \alpha')^{-1}$ as can 
be obtained from (\ref{weak_Mp}). 
Therefore we shall not be able to find a QCD-axion with a decay constant within the
preferred axion window.
In the strongly coupled heterotic M-theory the situation can be different. One 
might obtain axion decay constants much lower than the Planck scale 
as we are going to discuss in the next subsection.

\subsection{Heterotic M-theory}

Because of the clockwork parameter $m_\A$ present in \eqref{cw_axion}, the 4D zero modes of $\phi_L$ and $\phi_R$ are exponentially localized towards the left boundary $(x^{11}=0)$ and right boundary $(x^{11}=\pi r_{11})$, respectively. The 5D fields may be expanded in the 4D modes as
\begin{eqnarray}
\phi_L (x^\mu, x^{11}) &=& \sqrt{\frac{m_\A}{1-e^{-2m_\A  \pi r_{11}}}} e^{-m_\A x^{11}} a_L (x^\mu) + \sum_{n=1}^{\infty} \psi_L^{(n)}(x^{11}) \phi^{(n)}_L (x^\mu),  \label{phi_L}
\\
\phi_R (x^\mu, x^{11}) &=& \sqrt{\frac{m_\A}{e^{2m_\A \pi r_{11}}-1}} e^{m_\A x^{11}} a_R (x^\mu) + \sum_{n=1}^{\infty} \psi_R^{(n)}(x^{11}) \phi^{(n)}_R (x^\mu),  \label{phi_R}
\end{eqnarray}
where $a_L(x^\mu)$ and $a_R(x^\mu)$ denote 4D zero modes which will play the role of axions (with the prefactors chosen to normalize their 4D kinetic terms canonically), while $\phi^{(n)}_{L,R}(x^\mu)$ are 4D massive KK modes (with profiles in the 11-th dimension given by $\psi^{(n)}_{L,R}(x^{11})$).

As is obvious from \eqref{cw_axion}, axions $a_L$ and $a_R$ do not mix via their kinetic terms but both couple to $\tr F\widetilde{F}$ at each brane, because $a$ contributes to $\phi_L$ and $\phi_R$ as shown in \eqref{phiL} and \eqref{phiR}. The 4D Lagrangian for these axions has the form
\begin{eqnarray}
{\cal{L}}_{4,\rm axion}
=
&-&\frac{1}{2} (\partial_\mu a_L)^2-\frac{1}{2} (\partial_\mu a_R)^2 
\nonumber\\
&+& \frac{1}{16\pi^2}\left[\left(\frac{a_L}{f_{L1}}-\frac{a_R}{f_{R1}} \right){\rm tr} F_{(1)} \widetilde{F}_{(1)} +\left(\frac{a_L}{f_{L2}}- \frac{a_R}{f_{R2}} \right){\rm tr} F_{(2)} \widetilde{F}_{(2)} \right],
\label{4D_axion_LR}
\end{eqnarray}
where
\begin{align}
&\displaystyle f_{L1}=\frac{f_a}{\cos\beta}\sqrt{1-\exp(-2m_\A\pi r_{11})}
\,,
&\displaystyle f_{L2}=\frac{f_a}{\cos\beta}
\frac{\sqrt{\exp(2m_\A\pi r_{11})-1}}
{\exp\left((\frac12k_2-k_1)\pi r_{11}\right)}\,,
\\
&\displaystyle f_{R1}=\frac{f_a}{\sin\beta}\sqrt{\exp(2m_\A\pi r_{11})-1}\,,
&\displaystyle f_{R2}=\frac{f_a}{\sin\beta}
\frac{\sqrt{1-\exp(-2m_\A\pi r_{11})}}
{\exp\left((\frac12k_2-k_1)\pi r_{11}\right)}
\,,
\end{align}
and
\begin{equation}
f_a = \frac{F_a^{3/2}}{4\pi\sqrt{m_\A}} =\frac{\sqrt{3\pi}}{4\pi} \frac{\alpha_{\rm YM}^{1/6}}{\sqrt{|n_1|}} \frac{1}{\ell_{11}} \sim 0.1 M_{11} \,. \label{fa}
\end{equation}
The couplings of a given axion to two branes are very different for large values of $\pi r_{11}$ due to the clockwork mechanism:
\begin{eqnarray}
\frac{f_{L1}}{f_{L2}}
&=&
\exp\left[-\left(m_\A-\left(\frac12 k_2 - k_1\right)\right)\pi r_{11}\right]
\,,
\\
\frac{f_{R2}}{f_{R1}}
&=&
\exp\left[-\left(m_\A+\left(\frac12 k_2 - k_1\right)\right)\pi r_{11}\right]
\,.
\end{eqnarray}
The coupling of the axion $a_L$ ($a_R$) to the left (right) brane at which it is exponentially localized is much bigger than its coupling to the right (left) brane.

The above exponential factors can be expressed in terms of the 4D Planck scale $\Mp$ and the 11D string scale $M_{11}$. 
According to \cite{Im:2018dum},
\dis{
\frac{\Mp}{M_{11}} \simeq \gamma \,e^{k \pi r_{11}}\,, 
}
where the factor $\gamma = {\cal O}(10\div100)$ depends on compactification while $k \equiv  \frac12 k_2 + k_1$
is the clockwork parameter controlling the coupling hierarchy for the zero mode graviton.

Notice that this clockwork parameter is the same as the one for axions, i.e.~$m_\A = k$ 
as can be seen in (\ref{mA}). Therefore,
\begin{align}
\exp{(m_\A \pi r_{11})} &\simeq \frac{1}{\gamma} \frac{\Mp}{ M_{11}} \,,
\\
\exp{\left(\Big(\frac12k_2 - k_1\Big)\pi r_{11}\right)} &\simeq \left( \frac{1}{\gamma} \frac{\Mp}{ M_{11}} \right)^{(\hat{c}^2 -2)/(\hat{c}^2 +2 )},
\end{align}
where the exponent of the second term $(\hat{c}^2 -2)/(\hat{c}^2 +2 )=1/2$ for $\hat{c}^2 = 6$. 
This implies
\begin{align}
f_{L1} &\sim 0.1 M_{11}\,,
&f_{L2} \sim \frac{0.1}{\sqrt{\gamma}} \sqrt{ \frac{ M_{11}}{\Mp}}\,\Mp\,, 
\label{fL} 
\\
f_{R1} &\sim \frac{0.1}{\gamma} \Mp\,, 
&f_{R2} \sim 0.1 \sqrt{\gamma} \sqrt{ \frac{ M_{11}}{\Mp}}\,M_{11}  \,. 
\label{fR}
\end{align}

We see that in the heterotic M-theory case we could obtain axion decay constants that
are small compared to the Planck scale. This mechanism requires a rather large value of
$\pi r_{11}$ and thus a small value of the scale $M_{11}$. In fact, the above equations
apparently reveal several options. In the case of $f_{L1}$, for example, we have to choose $M_{11}$ to be an order of magnitude bigger than the axion decay constant.
The strongest suppression appears
for $f_{R2}$, where $M_{11}$ might even be $2\div3$ orders of magnitude above the axion window.
We might thus hope to obtain acceptable 
QCD-axion candidates for sufficiently low values of $M_{11}$. In the next section 
we shall see that among these various candidates there is only one, unique solution
to the strong CP-problem.

\section{Towards a QCD axion}
\label{sec_QCD_axion}

Let us finally turn to the search for a suitable  QCD axion candidate. It should have a small decay
constant and we thus assume throughout this section that $\pi r_{11}$ is sufficiently large. The axion 
should, of course, have a sizeable coupling to the observable brane (where QCD is 
located) and a potential that is flat enough to dynamically adjust $\theta_{\rm QCD}$ to zero.
The candidate should be a suitable combination of $a_L$ and $a_R$, as defined above.

The fields $a_L$ and $a_R$ are both massless at the level of the action \eqref{cw_axion} but become massive when the instanton effects are taken into account. Then, as we will see below, in general they are not mass eigenstates. The potential for the axions generated by instantons has the form
\begin{eqnarray}
\label{Vaxions}
V(a_L, a_R) = &-& \Lambda_{\rm QCD}^4 \cos \left( \frac{a_L}{f_{L1}} - \frac{a_R}{f_{R1}} + \theta_{\rm QCD} \right) 
- \sum_i \Lambda_{(1) i}^4 \cos \left( \frac{a_L}{f_{L1}} - \frac{a_R}{f_{R1}} + \theta_{(1) i} \right)  \nonumber\\
&-&\sum_j \Lambda_{(2) j}^4 \cos \left( \frac{a_L}{f_{L2}} - \frac{a_R}{f_{R2}} + \theta_{(2) j} \right) \, ,
\end{eqnarray}
where $\Lambda_{(1)i}^4$ and $\Lambda_{(2)j}^4$ are terms induced by instantons other than QCD from the visible sector
and hidden sector, respectively. The conditions for the minimum of this potential are:
\begin{equation}
\label{eom1}
 \Lambda_{\rm QCD}^4 \sin \left( \frac{a_L}{f_{L1}} - \frac{a_R}{f_{R1}} + \theta_{\rm QCD} \right) + \sum_i \Lambda_{(1) i}^4 \sin \left( \frac{a_L}{f_{L1}} - \frac{a_R}{f_{R1}} + \theta_{(1) i} \right) =0 \,, 
\end{equation}
\begin{equation}
\label{eom2}
\sum_j \Lambda_{(2) j}^4 \sin \left( \frac{a_L}{f_{L2}} - \frac{a_R}{f_{R2}} + \theta_{(2) j} \right)  =0 \,. 
\end{equation}
In order to solve the strong CP problem, the visible sector instantons other than the QCD 
contribution must be suppressed,
\begin{equation}
\label{strongCP}
\Lambda_{(1)i}^4 \lesssim 10^{-10} \Lambda_{\rm QCD}^4\,, 
\end{equation}
so in the following we set $\Lambda_{(1)i}=0$. 
The contributions to the potential from the hidden sector instantons, $\Lambda_{(2)j}^4$, may be sizable. For any values of $\Lambda_{\rm QCD}$ and $\Lambda_{(2)j}$ the fields $a_L$ and $a_R$ acquire such vacuum expectation values that both minimalization conditions 
\eqref{eom1} and \eqref{eom2} are fulfilled and $\theta_{\rm QCD}^{\rm\,eff}=0$.

The second derivatives of the potential \eqref{Vaxions} give the mass squared matrix for the axions
\begin{equation}
\label{m2JK}
\left(m^2\right)_{JK}
= 
\frac{\Lambda_{\rm QCD}^4}{f_{J1}f_{K1}}+\frac{\bar{\Lambda}^4}{f_{J2}f_{K2}}\,,
\end{equation}
where $J,K=L,R$ and $\bar{\Lambda}$ is defined by
\begin{equation}
\label{Lambda_bar}
\bar{\Lambda}^4
=
\left.\sum_j \Lambda_{(2) j}^4 \cos \left( \frac{a_L}{f_{L2}} - \frac{a_R}{f_{R2}} + \theta_{(2) j} \right)
\right|_{\sum_j \Lambda_{(2) j}^4 \sin \left( \frac{a_L}{f_{L2}} - \frac{a_R}{f_{R2}} + \theta_{(2) j} \right)=0}\,.
\end{equation}
The mass squared matrix \eqref{m2JK} has two eigenvalues
\begin{equation}
m^2_{a_h,a_l}
=
\frac{m^2_{LL}+m^2_{RR}}{2}
\pm
\frac{m^2_{LL}-m^2_{RR}}{2}
\sqrt{1+\frac{2m^2_{LR}}{m^2_{LL}-m^2_{RR}}}\,,
\end{equation} 
corresponding to the eigenvectors 
\begin{equation}
 \left(%
\begin{array}{c}
a_h \\
a_l
\end{array}%
\right)
=
\left(%
\begin{array}{cc}
\cos \beta' & \sin \beta' \\
-\sin \beta' & \cos \beta'
\end{array}%
\right)
 \left(%
\begin{array}{c}
 a_L \\
 a_R
\end{array}%
\right) \,.
\end{equation}
The rotation angle $\beta'$ fulfills the condition
\begin{equation}
\tan2\beta'
=
\frac{2m^2_{LR}}{m^2_{LL}-m^2_{RR}}\,.
\end{equation}
The axion action written in terms of the mass eigenstates has form analogous to \eqref{4D_axion_LR}:
\begin{eqnarray}
\label{4D_axion_hl}
{\cal{L}}_{4,\rm axion}
=
&-&\frac{1}{2} (\partial_\mu a_h)^2-\frac{1}{2} (\partial_\mu a_l)^2 
\nonumber\\
&+& \frac{1}{16\pi^2}\left[\left(\frac{a_h}{f_{h1}}-\frac{a_l}{f_{l1}} \right){\rm tr} F_{(1)} \widetilde{F}_{(1)} +\left(\frac{a_h}{f_{h2}}- \frac{a_l}{f_{l2}} \right){\rm tr} F_{(2)} \widetilde{F}_{(2)} \right],
\end{eqnarray}
where constants $f_{hi}$ and $f_{li}$ are related to $f_{Li}$ and $f_{Ri}$ by appropriate rotations by 
angle $\beta'$.

In general, both mass eigenstate axions, $a_h$ and $a_l$, couple to both branes and so both could contribute to the solution of the strong CP problem. However, the situation simplifies considerably in the 
 case of interest to us where a large $\pi r_{11}$ leads to $M_{11}\ll \Mp$. Such strong inequality is necessary, as follows from eqs.~\eqref{fL} and \eqref{fR}, to get any axion decay constants substantially smaller than the Planck scale, e.g.~in the axion window (and helps also to address the hierarchy problem). The parameters $\Lambda_{(2)j}^4$ may be written as
\dis{
\label{hidden_YM_inst}
\Lambda_{(2)j}^4 = \Lambda_{{\rm c}(2)j}^4 \, e^{-8\pi^2/g^2} \sim  \Lambda_{{\rm c}(2)j}^4 \, e^{-16\pi^2 \hat{V} }\,,
}
where $ \Lambda_{{\rm c}(2)j}$ is the confinement scale of the gauge field $F_{(2) j}$. The CY volume modulus $\hat{V}$ at the second boundary may be  estimated as
\dis{
\V = e^{k_2 \pi r_{11}} \sim \left(\frac{\Mp}{M_{11}} \right)^{3/2} \,.
}
Therefore, for $M_{11}\ll \Mp$, parameters $\Lambda_{(2)j}$ and their combination $\bar{\Lambda}^4$ defined in \eqref{Lambda_bar} will be typically very small. In the limit $\bar{\Lambda}^4\ll\Lambda_{\rm QCD}^4$ one may neglect the contribution from $\bar{\Lambda}^4$ to the axion mass matrix given by the second term on the rhs of \eqref{m2JK}. Then, $m_{a_h}^2\approx\Lambda^4_{\rm QCD}\sqrt{f_{L1}^{-2}+f_{R1}^{-2}}\gg m_{a_l}^2$ and the expression for the rotation angle $\beta'$ simplifies to $\tan\beta'=f_{L1}/f_{R1}$.
The couplings of the axions in this limit read
\begin{align}
&\frac{1}{f_{h1}}=\sqrt{\frac{1}{f_{L1}^2}+\frac{1}{f_{R1}^2}}
\,,
&\frac{1}{f_{h2}}=\frac{1}{\sqrt{f_{L1}^2+f_{R1}^2}}
\left(\frac{f_{R1}}{f_{L2}}+\frac{f_{L1}}{f_{R2}}\right)
\,,\\
&\frac{1}{f_{l1}}\sim 0
\,,
&\frac{1}{f_{l2}}=\frac{1}{\sqrt{f_{L1}^2+f_{R1}^2}}
\left(\frac{f_{R1}}{f_{R2}}-\frac{f_{L1}}{f_{L2}}\right)
\,.
\end{align}
The lighter axion, $a_l$, essentially decouples from QCD. The heavier axion, $a_h$, solves the strong CP problem but couples also to the hidden brane ($f_{h2}\ne\infty$). 
From (\ref{fL}) and (\ref{fR}), the decay constants are estimated as
\bea
f_{h1} \sim 0.1 M_{11}\,, \quad \quad f_{h2} \sim 0.1 \gamma^{-1/2} \sqrt{\Mp M_{11}} ,
~\quad f_{l2} \sim 0.1 \gamma^{1/2} M_{11} \sqrt{\frac{M_{11}}{\Mp}} \label{fhfl}
\eea
Thus our QCD axion $a_h$ has the decay constant about one order of magnitude below the string scale $f_{h1} \sim 0.1 M_{11}$. 
So, if the string scale $M_{11}$ lies 
in the ``axion window" moved up by one order of magnitude, 
\dis{
10^{10}\, {\rm GeV} \lesssim M_{11} \lesssim 10^{13} \,{\rm GeV}\,, \label{QCD_M11}
} 
a phenomenologically acceptable solution to the strong CP problem is realized. 
In our previous work \cite{Im:2018dum}, we showed that the string scale $M_{11}$ can be lowered 
down  to the TeV scale through a large 
warped 11-th dimension without violating current experimental bounds on the size of extra dimension. 
So it is not difficult to have the string scale 
lying in or slightly above the axion window as a result of the warping.

The relevant candidate for the QCD-axion is thus $a_{h}$. There is a second axion $a_{l}$
whose decay constant $f_{l2}$ is parametrically even smaller than $M_{11}$. However, it
cannot play the role of a QCD-axion as it essentially decouples from the observable sector.
This observation extends and completes previous discussions of axions embedded in string
theory \cite{Svrcek:2006yi}.

The second axionic state $a_{l}$ couples predominantly to the hidden sector. In the case of
large $r_{11}$ under consideration, its mass will be extremely small as the contributions of
the hidden sector instantons are tiny. The appearance of the ultra-light axion is a direct
consequence of the presence of the QCD-axion $a_{h}$ with decay constant compatible
with the ``axion window". The mass of the ULA, its phenomenological
properties and possible cosmological implications
\cite{Hlozek:2014lca}
are strongly model
dependent and should be subject to future investigations.

\section{Conclusions and Outlook}
\label{sec_conclusions}

In the present paper we have applied the previously discussed clockwork mechanism
within the framework of heterotic M-theory  \cite{Im:2018dum} to axionic fields that could
serve as candidates for a solution to the strong CP-problem. The decay constant $f_a$ of the
corresponding axion field should be inside the ``axion window" around $10^{11}$ GeV.
Although there is a variety of axion fields in the 10D heterotic string theory, it is impossible 
to fulfill this requirement. In 11D heterotic M-theory, however, this is different. We can lower
the axion decay constants under the same conditions and for the same reasons as we lowered
the weak scale to the TeV-scale in ref.~\cite{Im:2018dum}. The typical set-up is a warped 
7D compactification as the semi-direct
product of an interval (with length $\pi r_{11}$) and a 
6D Calabi-Yau manifold. A lower value of $f_a$ requires a rather large value of $r_{11}$
and a smaller value of $M_{11}$ (the fundamental mass scale of M-theory). At an intermediate 
step in this compactification, the candidate 4D axion fields will reside in 
5D supergravity multiplets. In a hypermultiplet the 4D axion derives from a 5D scalar field,
whereas in a gravity-multiplet and a vector-multiplet (relevant if  $\hone>1$) it is a component of a 5D vector field. Generically, the
4D axion candidate will be a linear combination of these multiplets. As a prototype model
we consider a Calabi-Yau manifold with $\hone =1$ and thus a two-axion system with
one of these fields each. The case for $\hone > 1$ is qualitatively similar and is discussed
in appendix A.

A priori we can find various values for the axion decay constants under consideration
(see e.g.~formulae (4.15-16)). Low values of $f_a$ are possible in the case of large
$r_{11}$ and small $M_{11}$. If we concentrate on an axion that could solve the strong
CP-problem, however, only one of these choices is possible: the one with
the string scale one order of magnitude above the axion decay constant: $M_{11}\sim10f_a$. 
This QCD-axion is accompanied by a
second axion that is essentially decoupled form the observable sector gauge
groups. Due to a type of ``see-saw" mechanism, this second axion is
extremely light (due to the suppression of the hidden sector instanton contributions).
The decay constant of the ultra-light axion (ULA) is smaller than that of the QCD-axion
(as $f_{l2}$ in formula (5.15)).
The exact value of the mass and the phenomenological properties of the ULA are
strongly model dependent (especially in the case $\hone > 1$). Although the ULA
decouples from the observable sector and will be difficult to detect in particle physics
experiments, the stronger coupling to the hidden sector might eventually be of
cosmological relevance. This should be subject to future investigations.

\acknowledgments

SHI acknowledges partial support from Basic Science Research Program through the National Research Foundation of Korea (NRF) funded by the Ministry of Education (NRF-2017R1D1A1B06033701 and 2019R1I1A1A01060680) and the Korea government (MSIP) (NRF-2018R1C1B3001379). 
SHI also thanks the CERN-Korea TH Institute, 
where valuable comments were given to this work. 
HPN thanks the CERN Theory Department for hospitality and support.
MO acknowledges partial support from National Science Centre, Poland, 
grants DEC-2015/18/M/ST2/00054 and DEC-2016/23/G/ST2/04301.

\appendix
\section{Axion couplings for $\mathbf{\hone=2}$}
\label{app_h11=2}
In this Appendix, we provide more detailed formulae for the axion lagrangian and work out
a simple example for $\hone > 1$. The 5D axion effective lagrangian is given in (\ref{5D_axion}):
\begin{align}
\S_{5, {\rm axion}}=&\, -2\pi \int_{\M^5} \left[
\frac{\lh^3}{2 V_X} 
\left( \d a + {n_i \A^i}\right) \wedge \star \left( \d a + {n_i \A^i}\right) 
+ \frac{1}{2\lh^3} G_{ij}\,\d\A^i \wedge \star \d\A^j 
\right]
\nonumber\\
 &\, + \int_{\M^5} \frac{1}{4\pi}\, a 
\left[
\left( {\rm tr} F_{(1)} \wedge F_{(1)} \right)\delta(x^{11}) 
+ \left( {\rm tr} F_{(2)} \wedge F_{(2)} \right) 
\delta(x^{11}-\pi r_{11})
\right]
\wedge\d x^{11} \,.
\end{align}
The kinetic matrix $G_{ij}$ for the gauge fields is given by (\ref{Gij}):
\dis{ \label{A_Gij}
G_{ij} &= \int_{\X^6} \omega_i^{(1,1)} \wedge \star \omega_j^{(1,1)} \\
&=  -V_X^{1/3} \left[ d_{ijk} X^k - \frac14 (d_{ilm} X^l X^m) (d_{j n p} X^n X^p) \right]
}
where the indices run over $i,j,k,\ldots =1, \dots, \hone$ and 
\dis{
d_{ijk} \equiv \int_{\X^6} \omega_i^{(1,1)} \wedge \omega_j^{(1,1)}  \wedge \omega_k^{(1,1)} 
}
are the CY intersection numbers, and the K\"ahler moduli $X^i$ satisfy the constraint
\dis{ \label{Kahler_crt}
\frac{1}{6} d_{ijk} X^i X^j X^k = 1.
}
In the case of $\hone=1$, $d_{111} = 6$ and $X^1 = 1$ so that $G_{11} = 3V_X^{1/3}$ as used in (\ref{G11}).

As a simple example with $\hone > 1$, let us consider $\hone=2$ and the CY intersection number $d_{112}\neq0$ while the other 
$d_{ijk}$ vanish. This example was also considered in \cite{Im:2018dum}.
In this case, the constraint (\ref{Kahler_crt}) becomes
\dis{
d_{112} (X^1)^2 X^2 = 2,
}
which may be solved by
\dis{
X^1 &= \frac{1}{\eta} e^{-bS_1}, \\
X^2 &= \frac{1}{\eta} e^{2b S_1},
}
where $\eta = (d_{112}/2)^{1/3}$ and $b=1/\sqrt3$ to canonically normalize $S_1$. 
On the other hand, the kinetic matrix (\ref{A_Gij}) comes out as
\dis{ \label{Gsol}
G_{11} &= 2V_X^{1/3} (X^1)^{-2}, \\
G_{22} &= V_X^{1/3} (X^2)^{-2}, \\
G_{12} &= 0.
}

It was shown in \cite{Im:2018dum} that there are three qualitatively different background solutions. 
In the first case, for which both flux numbers $n_1$ and $n_2$ are non-vanishing\footnote{Here the flux number $n_i$ is related to the dimensionful flux factor $\mu_i$ defined in \cite{Im:2018dum}
by the following relation,
\dis{
n_i = - \frac{V_{X,0}^{2/3}}{4\sqrt{2} \, \lh^3} \mu_i
}
},
the K\"ahler modulus $S_1$ is stabilized at $\sqrt{3} S_1 = \ln (n_1/2n_2)$, and the 
background geometry is the same as for $\hone=1$, i.e.~as given in (\ref{h11=1_bc}) and (\ref{h11=1_V}). 
The gauge couplings for $\A_1$ and $\A_2$ are then
\dis{
g_{\A_1}^2 &= \frac{\lh^3}{6\pi V_{X,0}^{1/3}} \frac{3}{2\eta^2} \left( \frac{2n_2}{n_1}\right)^{2/3}, \\
g_{\A_2}^2 &= \frac{\lh^3}{6\pi V_{X,0}^{1/3}} \frac{3}{\eta^2} \left( \frac{n_1}{2n_2}\right)^{4/3}.
}
Compared to $\hone=1$, we just have to make the following replacement 
in the gauge field basis where the gauge couplings do not appear in kinetic terms:
\dis{
\A &\,\rightarrow\, {\tilde{\A}}\equiv\frac{n_1 g_{\A_1} \A_1 + n_2 g_{\A_2} \A_2}{\sqrt{n_1^2 g_{\A_1}^2 + n_2^2 g_{\A_2}^2}}\,, \\
n_1 g_{\A} &\,\rightarrow\, \sqrt{n_1^2 g_{\A_1}^2 + n_2^2 g_{\A_2}^2}\,.
}
Only the above linear combination of the gauge fields mixes with $a$, while the orthogonal one does not. So the gauge field orthogonal to $\tilde{\A}$ is irrelevant for the axion lagrangian unless there is another
source of axion mixing.\footnote{Including membrane instanton effects (e.g.~\cite{Harvey:1999as}), the additional axion candidate 
from the 11-th component of this orthogonal gauge field can mix with $\tilde\A_{11}$. However, this effect is strongly model-dependent and typically small as the membrane instanton amplitude is either comparable to or smaller than the hidden YM instanton amplitude (\ref{hidden_YM_inst}).} 
Then it turns out that the clockwork effective lagrangian is equivalent to (\ref{cw_axion}) with the same 
gauge boson mass $m_\A$, and the clockwork fields $\phi_L$ and $\phi_R$ are related to $a$ and $\tilde{\A}_{11}$
by the same relation (\ref{phiL})-(\ref{phiR}) and mixing angle (\ref{mixing_angle}).

The two other cases are obtained when one of the flux numbers is vanishing, i.e.~either $n_1=0$ or $n_2=0$. 
For these cases, obviously only one of the gauge fields ($\A_2$ or $\A_1$, respectively) mixes with $a$, so the
other gauge field does not participate in the axion lagrangian at this level.
It was shown in \cite{Im:2018dum} that for these cases the K\"ahler modulus $S_1$ is  identified as a component
of the GLD dilaton together with the volume modulus $S_V = (1/\sqrt{2}) \ln \hat{V}_X$.
This renders the geometric parameters $\hat{b}$ and $\hat{c}^2$ in (\ref{k1k2}) to be different from the $\hone=1$ case. 
In particular, $\c^2=k_2/k_1$, which is 6 in the case of $\hone=1$, becomes
\dis{
\c^2 = \begin{cases} 7 & {\rm for}\,\,\, n_1\ne0,\,\,n_2=0 
\\ 10 & {\rm for}\,\,\, n_1=0,\,\, n_2\ne0 \end{cases}\,.
}
Furthermore, since $S_1$ has a $x^{11}$-dependent background value, the gauge fixing term (\ref{Sgf}) should be generalized by the replacement
\dis{
V_X^{1/3} \,\rightarrow\, V_X^{1/3} \exp\left(2 \left(1-\frac{6}{\c^2}\right) k_2 x^{11}\right).
} 
The resultant clockwork effective lagrangian still has the same form as (\ref{cw_axion}). The gauge boson mass $m_\A$ turns out to be 
\dis{
m_\A = \sqrt{\left(\frac12k_2 -k_1 - \left(1-\frac{6}{\c^2}\right) k_2 \right)^2 + n_i^2 g_\A^2 F_a^3} = \frac12 k_2 + k_1\,,
}
so it has the same form as (\ref{mA}). On the other hand, the clockwork fields $\phi_L$ and $\phi_R$ are related to $a$ and $\A_{11}$
via the generalized relations
\bea
\label{phiL_gen}
\phi_L 
&=&
e^{-\left(\frac12 k_2-k_1- \left(1-\frac{6}{\c^2}\right) k_2\right) x^{11}}
\left(a\,F_a^{3/2} \,\cos\beta
-
\A_{11}\,g_\A^{-1}\, \sin\beta\right),  
\\
\label{phiR_gen}
\phi_R
&=&
e^{-\left(\frac12 k_2-k_1- \left(1-\frac{6}{\c^2}\right) k_2\right) x^{11}}
\left(-a\,F_a^{3/2} \,\sin\beta
-
\A_{11}\,g_\A^{-1}\,\cos\beta\right), 
\eea
with the mixing angle $\beta$ satisfying the condition
\begin{equation}
\tan 2\beta = -g_\A F_a^{3/2} \frac{n_i}{\frac12k_2-k_1- \left(1-\frac{6}{\c^2}\right) k_2}
=\frac{\sqrt{\left(2 k_1+ \left(1-\frac{6}{\c^2}\right) k_2\right)\left(k_2 -\left(1-\frac{6}{\c^2}\right) k_2\right)}}{\frac{1}{2} k_2-k_1- \left(1-\frac{6}{\c^2}\right) k_2} \,. \label{mixing_angle_gen}
\end{equation}

\section{Kaluza-Klein axions} 
\label{KK_axions}

In this section we discuss properties of Kaluza-Klein axion excitations in heterotic M-theory.
Their spectrum and couplings can be obtained from the continuum clockwork lagrangian (\ref{cw_axion}).
Detailed calculations were carried out in \cite{Choi:2017ncj}. The analysis indicates that both $n$-th KK states, $\phi_L^{(n)}$ and $\phi_R^{(n)}$ with $n=1, 2, \dots$, have the same mass which is approximately given by
\dis{
M_n \approx \left(n-\frac{1}{4} + \frac{m_\A}{2|p|} \right) \pi |p| \exp(-|p|\pi r_{11})\,,
}
where $p \equiv k_1 - k_2$, 
as defined in \cite{Im:2018dum}, controls the clockwork gear (KK mode) masses.
 In terms of the 11D Planck mass $M_{11}$ and 4D Planck mass $\Mp$, this is estimated as   \cite{Im:2018dum} 
\dis{
M_n \sim 30\, n  M_{11} \left( \frac{M_{11}}{\Mp} \right)^{2(\c^2-1)/(\c^2+2)}\,.
}
Applying the heterotic bound $\c^2 \geq 6$ argued in  \cite{Im:2018dum},  
the lightest KK mode mass $M_1$ is parametrically smaller than $M_{11}$.
However, heavy KK modes with very large $n$ can have masses comparable to $M_{11}$ for which
the above approximate formula is still valid.
 
The couplings to the boundary fields may be written as
\dis{
\frac{1}{16\pi^2}\left[\sum_{n=1}^\infty \left( \frac{\phi_L^{(n)}}{f_{L1}^{(n)}} -  \frac{\phi_R^{(n)}}{f_{R1}^{(n)}}\right){\rm tr} F_{(1)} \widetilde{F}_{(1)} + \sum_{n=1}^\infty \left( \frac{\phi_L^{(n)}}{f_{L2}^{(n)}} -  \frac{\phi_R^{(n)}}{f_{R2}^{(n)}}\right){\rm tr} F_{(2)} \widetilde{F}_{(2)} \right] \,,
}
where 
\begin{align}
&\displaystyle f_{L1}^{(n)}\approx\frac{f_a}{\cos\beta} \sqrt{\frac{m_\A}{|p| \pi}} \,\Gamma\left(\frac{m_\A}{|p|}\right) \left[ \frac{\pi}{2} \left( n-\frac{1}{4} + \frac{m_\A}{2|p|}\right)\right]^{-\frac{m_\A}{|p|}+\frac{1}{2}} \exp(m_\A \pi r_{11})
\,,\\
&\displaystyle f_{L2}^{(n)}\approx (-1)^n \frac{f_a}{\cos\beta} \sqrt{\frac{m_\A}{|p| \pi}} \exp\Big(-\big(\frac12k_2-k_1\big) \pi r_{11}\Big)\,,\\
&\displaystyle f_{Ri}^{(n)}\approx\cot\beta f_{Li}^{(n)}\,,
\end{align}
with $f_a$ defined in (\ref{fa}). 
This means
\bea
f_{L1}^{(n)} &\sim& f_{R1}^{(n)} \sim \Mp \left( \frac{M_1}{M_n}\right)^{3/2(\c^2-1)}\,, \\
f_{L2}^{(n)} &\sim& f_{R2}^{(n)} \sim M_{11} \left(\frac{M_{11}}{\Mp}\right)^{(\c^2-2)/(\c^2+2)}\,.
\eea
We see that light KK modes with small $n$ couple to the visible sector with decay constants around the 4D Planck scale $\Mp$. 
Heavy KK modes with $M_n \sim M_{11}$ can have quite lower decay constants below $\Mp$ as pointed out in \cite{Im:2018dum}
as a characteristic property of the heterotic clockwork due to $\c^2 > 1$.
On the other hand, the decay constants $f_{L2}^{(n)}$ and $f_{R2}^{(n)}$ to the hidden sector are parametrically smaller than $M_{11}$.
Nevertheless, since the hidden sector gauge coupling is small when $M_{11}$ is below $\Mp$,
\dis{
g_{(2)}^2 \sim \hat{V}_X^{-1} \sim \left( \frac{M_{11}}{\Mp}\right)^{12/(\c^2+2)}\,,
}  
the net coupling to the hidden sector is
\dis{
\frac{g_{(2)}^2}{f_{L2}^{(n)}} \sim \frac{g_{(2)}^2}{f_{R2}^{(n)}} \sim \frac{1}{M_{11}} \left(\frac{M_{11}}{\Mp}\right)^{(14-\c^2)/(\c^2+2)}\,.
}
This is still around $1/\Mp$ or quite smaller than
$1/M_{11}$ for the known examples $\c^2 = 6, 7, 10$ in \cite{Im:2018dum}.

Let us focus on the simplest example $\c^2 = 6$ with $M_{11}$ lying around the axion window $(10^{10} \,{\rm GeV} < M_{11} < 10^{13} \,{\rm GeV})$ 
in order to solve the strong CP problem with the zero mode axion. 
The KK axion masses are then
\dis{ \label{KK_mass}
M_n \approx 9\,n \,{\rm GeV} \,\left(\frac{M_{11}}{10^{10} \,{\rm GeV}} \right)^{9/4} \,,
} 
while the KK axion decay constants are
\bea
f_{L1}^{(n)} &\sim& f_{R1}^{(n)} \sim \frac{\Mp}{n^{3/10}} > 2.4 \times 10^{15} {\rm GeV} \left(\frac{M_{11}}{10^{10}\, {\rm GeV}} \right)^{3/8}\,, \\
f_{L2}^{(n)}/g_{(2)}^2 &\sim& f_{R2}^{(n)}/g_{(2)}^2 \sim \Mp\,,
\eea
where the net couplings including the gauge coupling are estimated for the hidden sector because the gauge coupling may be small.
Therefore, if $M_{11} = 10^{10}$ GeV for instance, we have KK axions of 10 GeV scale mass gap with decay constants ranging from the 4D Planck scale all the way down to
about $10^{15}$ GeV.
These are so-called axion-like particles (ALP) present in heterotic M-theory accompanying the QCD axion.

Typically the KK axions of masses below $10^6$ GeV have long enough life time to cause a problem for the standard cosmology.
A straightforward way to cure this problem is to have a Hubble scale during inflation lower than the lightest KK axion mass in order to
suppress their production \cite{Arvanitaki:2009fg}. As can be seen in (\ref{KK_mass}), the lightest KK axion mass $M_1$ for $M_{11}>10^{10}$ GeV is well above $H_{\rm inf} \sim 0.1$ GeV for successful baryogenesis.

\bibliography{heterotic_axion_v2}
\bibliographystyle{utphys}

\end{document}